\documentstyle[aps,prd]{revtex}
\begin{document}
\title{Hamiltonian analysis of General relativity with the Immirzi 
parameter}
\author{N. Barros e S\'a\thanks{Email address: 
nunosa@vanosf.physto.se. Supported by grant PRODEP-Ac\c c\~ao 5.2.}}
\address{Fysikum, Stockholms Universitet, Box 6730, 
113 85 Stockholm, Sverige\\
{\rm and} DCTD, Universidade dos A\c cores, 
9500 Ponta Delgada, Portugal}
\date{\today}
\maketitle

\begin{abstract}
Starting from a Lagrangian we perform the full 
constraint analysis of the Hamiltonian for General relativity in 
the tetrad-connection formulation for an 
arbitrary value of the Immirzi parameter and solve the second 
class constraints, presenting the theory with a Hamiltonian 
composed of first class constraints which are the generators of the 
gauge symmetries of the action. In the time gauge we then recover 
Barbero's formulation of gravity.
\end{abstract}

A formulation of General relativity using real connections as 
the dynamical variables of the theory has been proposed by 
Barbero \cite{bar} and has been used since then in the loop 
approach to quantum gravity \cite{rov}. 
The reality of the theory, as opposed to the 
complex variables of Ashtekar \cite{ash}, is achieved at the cost of 
the non-polynomiallity of the Hamiltonian constraint. 
Barbero's formulation also leads to the so-called Immirzi 
ambiguity \cite{imm}, which from the present point of view 
\cite{hol} arises from the addition to the standard 
action for General relativity in the tetrad-connection 
formalism of a term which does not affect the classical 
equations of motion, 
but which may affect the quantum theory \cite{rth}. The 
spectrum of the volume and area operators, and consequently the 
entropy of black holes, seems to depend on the 
Immirzi parameter $\beta$ (the constant that multiplies the term 
added to the the action that we mentioned before). 
Notably there is one single value of this 
parameter for which the conventional expression of the entropy of 
black holes is reproduced\cite{as2}. 
Ashtekar's gravity can be obtained 
from the complex version of the ordinary tetrad-connection action 
for gravity supplemented with the Immirzi term when 
$\beta=\pm i$ \cite{jsm,sam}, 
and Barbero's gravity can be obtained in the same fashion but with 
no need to complexify the theory and setting $\beta$ real and 
non-zero \cite{hol}. It seems therefore worthwhile to make an 
effort to 
understand better these theories. Here we perform the full 
Hamiltonian analysis prior to any gauge-fixing for an 
arbitrary value of the Immirzi parameter and solve all second 
class constraints, presenting the theory with a Hamiltonian 
composed of first class constraints which are the generators of the 
gauge symmetries of the action. In this way we are also able to 
confirm Holst's result \cite{hol} which has been called in question  
because it was obtained from a partially gauge-fixed action. 

In section \ref{se1} we introduce the reader to the Lagrangian of 
the theory and to the notation used. In section \ref{se2} the 
Hamiltonian is derived and in section \ref{seci} we compute the 
constraint algebra and find the secondary constraints. It turns 
out that some constraints form second class pairs. This analysis 
is similar to what is usually done without the Immirzi parameter, 
except that now there are two distinct natural choices of canonical 
variables. In section \ref{se4} the second class constraints are 
solved and the explicit form of the remaining first class 
constraints is presented. Finally in section \ref{se5} we fix the 
boost gauge freedom and show the equivalence of this theory to 
Barbero's formulation of gravity.

\section{The Lagrangian}\label{se1}

We use the standard tetrad field $e^{\alpha I}(x)$ and antisymmetric 
connection $\omega_{\alpha IJ}(x)$, both defined on each 
point $x$ of a spacetime manifold. We use Greek letters for indices 
in the tangent space to the manifold. Capital Latin letters stand 
for indices in the internal space, which is endowed with a 
Minkowskian metric with signature $(-1,1,1,1)$.
When $3+1$ 
decompositions will be performed, timelike indices will be labeled 
``t'' in the tangent space and ``0'' in the internal space, while 
spacelike indices will be labelled with small Latin letters 
starting at ``a'' and ``i'' for the tangent space and internal space 
respectively. When contracting indices in the internal space we 
shall not care about their position, whether raised or lowered, the 
presence of a suitable metric (Minkowskian in the whole space or 
Euclidean in the spacelike part of the decomposition) being 
understood when necessary.

The metric on the spacetime manifold is
\begin{equation}
g^{\alpha\beta}=e^{\alpha I}e^{\beta I}\ .
\end{equation}
The determinant of the tetrad field and the curvature in internal 
space are represented respectively by
\begin{eqnarray}
e&=&1/\det(e^{\alpha I})\\
R_{\alpha\beta IJ}&=&\partial_{[\alpha}\omega_{\beta ]IJ}+
\omega_{[\alpha|IK}\omega_{|\beta]JK}\ .
\end{eqnarray}
The dual of an antisymmetric tensor of second order in internal 
space is defined by
\begin{equation}
{^*U}_{IJ}=\frac{1}{2}\epsilon_{IJKL}U_{KL}\ .
\end{equation}

The action is a first order action constructed from the tetrads 
$e^{\alpha I}$ and connection $\omega_{\alpha IJ}$. It is 
$S=\int d^4x L$, with 
\begin{equation}
L=\frac{1}{2}ee^{\alpha I}e^{\beta J}\left( R_{\alpha\beta IJ}-
\frac{1}{\beta}{^*R_{\alpha\beta IJ}}\right) \ ,\label{lag}
\end{equation}
where $\beta$ is the Immirzi parameter, and it can be easily shown 
that it indeed describes locally General relativity at the classical 
level regardless of the value of $\beta$. 
It is very convenient to introduce for antisymmetric tensors 
of second order in internal space the notation 
\footnote{See the added note.} 
\begin{equation}
\stackrel{(\beta)}{U}_{IJ}=U_{IJ}-\frac{1}{\beta}{^*U_{IJ}}\ .
\label{fre1}
\end{equation}
This is a one-to-one relation between the $\beta$-dependent 
${^{(\beta)}U}_{IJ}$ and 
the $\beta$-independent $U_{IJ}$, with inverse
\begin{equation}
U_{IJ}=\frac{\beta^2}{1+\beta^2}
\left(\stackrel{(\beta)}{U}_{IJ}+
\frac{1}{\beta}{^*\stackrel{(\beta)}{U}_{IJ}}\right)\ .\label{fre2}
\end{equation} 
One can derive the useful relations 
\begin{eqnarray}
\stackrel{(\beta)}{(U_{[I|K}V_{|J]K})}&=&\stackrel{(\beta)}
{U}_{[I|K}V_{|J]K}=U_{[I|K}\stackrel{(\beta)}{V}_{|J]K}\\
\stackrel{(\beta)}{U}_{IJ}V_{IJ}&=&U_{IJ}
\stackrel{(\beta)}{V}_{IJ}\ .
\end{eqnarray}
With this notation the Lagrangian (\ref{lag}) becomes
\begin{equation}
L=\frac{1}{2}ee^{\alpha I}e^{\beta J}
\stackrel{(\beta)}{R}_{\alpha\beta IJ}\ .\label{lag22}
\end{equation}

\section{The Hamiltonian}\label{se2}

We perform the 3+1 decomposition of (\ref{lag22}),
\begin{equation}
L=ee^{tI}e^{aJ}\stackrel{(\beta)}{R}_{taIJ}+
\frac{1}{2}ee^{aI}e^{bJ}\stackrel{(\beta)}{R}_{abIJ}\ ,\label{tmu}
\end{equation}
and split the tetrad field into 
\begin{eqnarray}
N&=&-\frac{1}{eg^{tt}}\\
N^a&=&-\frac{g^{ta}}{g^{tt}}\\
\pi^{aIJ}&=&ee^{t[I|}e^{a|J]}\ .\label{mom}
\end{eqnarray}
The first component is the lapse and the second three are the 
shift. The $\pi^{aIJ}$ have 18 components. Since the tetrad has 
got 16 independent components, there are 6 variables in excess 
in this decomposition. Indeed the $\pi^{aIJ}$ are subject to 6 
constraints,
\begin{equation}
{\cal C}^{ab}=\frac{1}{2}\pi^{aIJ}{^*\pi^{bIJ}}\sim 0
\label{con}
\end{equation}

It is standard to derive that the second parcel on the right hand 
side of (\ref{tmu}) becomes
\begin{equation}
\frac{1}{2}ee^{aI}e^{bJ}\stackrel{(\beta)}{R}_{abIJ}=
N{\cal H}+N^a{\cal H}_a\ ,\label{tmu2}
\end{equation}
with
\begin{eqnarray}
{\cal H}_a&=&\frac{1}{2}\pi^{bIJ}
\stackrel{(\beta)}{R}_{abIJ}=\frac{1}{2}\stackrel{(\beta)}{\pi}
{^{bIJ}}
R_{abIJ}\label{vec}\\
{\cal H}&=&\frac{1}{2}\pi^{aIK}\pi^{bJK}\stackrel{(\beta)}
{R}_{abIJ}=\frac{1}{2}\stackrel{(\beta)}{\pi}{^{aIK}}\pi^{bJK}
R_{abIJ}\ .\label{ham}
\end{eqnarray}
And the first parcel on the right hand 
side of (\ref{tmu}) becomes
\begin{equation}
\frac{1}{2}\pi^{aIJ}\stackrel{(\beta)}{R}_{taIJ}=
\frac{1}{2}\stackrel{(\beta)}{\pi}{^{aIJ}}R_{taIJ}=
\frac{1}{2}\stackrel{(\beta)}{\pi}{^{aIJ}}\dot\omega_{aIJ}-
\frac{1}{2}\Lambda_{IJ}{\cal G}^{IJ}\ ,\label{tmu3}
\end{equation}
where a partial integration has been performed in the second 
equality, and 
\begin{eqnarray}
\Lambda_{IJ}&=&-\omega_{tIJ}\\
{\cal G}^{IJ}&=&D_a\stackrel{(\beta)}{\pi}{^{aIJ}}\ ,\label{gau}
\end{eqnarray}
In this last formula covariant differentiation is acting in the 
internal space only. Equation (\ref{gau}) reads explicitly
\begin{equation}
{\cal G}^{IJ}=\partial_a\stackrel{(\beta)}{\pi}{^{aIJ}}+
\omega_a{^{IK}}\stackrel{(\beta)}{\pi}{^{aJK}}-
\omega_a{^{JK}}\stackrel{(\beta)}{\pi}{^{aIK}}=
\partial_a\stackrel{(\beta)}{\pi}{^{aIJ}}+
\stackrel{(\beta)}{\omega}_a{^{IK}}\pi^{aJK}-
\stackrel{(\beta)}{\omega}_a{^{JK}}\pi^{aIK}\ .
\end{equation}

Putting (\ref{tmu}), (\ref{tmu2}) and (\ref{tmu3}) together, one 
ends up with the Hamiltonian
\begin{equation}
H=N{\cal H}+N^a{\cal H}_a+\frac{1}{2}\Lambda_{IJ}{\cal G}^{IJ}+
\frac{1}{2}c_{ab}{\cal C}^{ab}\ ,\label{hami}
\end{equation}
where (\ref{con}) arose as primary constraints resulting from 
the very definition (\ref{mom}) of the variables $\pi^{aIJ}$. 
The symplectic form is 
\begin{equation}
\frac{1}{2}\stackrel{(\beta)}{\pi}{^{aIJ}}\dot\omega_{aIJ}=
\frac{1}{2}\pi^{aIJ}\dot{\stackrel{(\beta)}{\omega}_{aIJ}}\ .
\label{sifo}
\end{equation}
There are therefore two sets of variables which are canonically 
conjugate and that can be naturally chosen to parameterize phase 
space, $\beta$-dependent connections together with 
$\beta$-independent momenta, or vice-versa,
\begin{equation}
\left\{ \stackrel{(\beta)}{\omega}_{aIJ}(x),\pi^{bKL}(y)\right\} =
\left\{ \omega_{aIJ}(x),\stackrel{(\beta)}{\pi}{^{bKL}}(y)\right\} =
\delta_a^b(\delta_I^K\delta_J^L-\delta_I^L\delta_J^K)
\delta^3(x-y)\ .\label{pbu}
\end{equation}
Both connections are $SO(3,1)$ connections. 

Thus our system is described by any of the canonical pairs 
(\ref{pbu}) 
and the totally constrained Hamiltonian (\ref{hami}), where 
$\omega_{tIJ}$, $N^a$ and $N$ play the role of Lagrange multipliers 
for respectively the constraints (\ref{gau}), (\ref{vec}) and 
(\ref{ham}), known as the gauge, vector and scalar constraints. 
When performing the constraint analysis one should use the 
inverting equations between $\beta$-dependent and 
$\beta$-independent quantities, equations 
(\ref{fre1})-(\ref{fre2}), in order to write the constraints 
solely in terms of canonical variables, whatever the set one 
chooses to use.
In order to facilitate the use of $\beta$-dependent connections, 
we derive 
\begin{equation}
\stackrel{(\beta)}{R}_{abIJ}=
\partial_{[\alpha}\stackrel{(\beta)}{\omega}_{\beta ]IJ}
+\frac{\beta^2}{1+\beta^2}
\left[
\stackrel{(\beta)}{\omega}_{[\alpha|IK}
\stackrel{(\beta)}{\omega}_{|\beta]JK}+
\frac{1}{\beta}{^*\left(\stackrel{(\beta)}{\omega}_{[\alpha|IK}
\stackrel{(\beta)}{\omega}_{|\beta]JK}\right)}\right]\ .
\end{equation}

\section{Constraint analysis}\label{seci}

In this section we perform the constraint analysis of the 
Hamiltonian derived in the previous section. The resulting algebra 
is independent of the set of canonical pairs chosen to perform the 
calculations. 

It is easy to check that the constraints ${\cal G}^{IJ}$ are the 
generators of internal gauge transformation, and that the 
combinations
\begin{equation}
\tilde{\cal H}_a={\cal H}_a-\frac{1}{2}\omega_{aIJ}{\cal G}^{IJ}=
\frac{1}{2}\left[ \pi^{bIJ}\partial_a
\stackrel{(\beta)}{\omega}_{bIJ}-
\partial_b\left(\stackrel{(\beta)}{\omega}_{aIJ}
\pi^{bIJ}\right)\right]=
\frac{1}{2}\left[ \stackrel{(\beta)}{\pi}{^{bIJ}}\partial_a
\omega_{bIJ}-\partial_b\left(\omega_{aIJ}
\stackrel{(\beta)}{\pi}{^{bIJ}}\right)\right]
\end{equation}
are the generators of spatial diffeomorphisms. It is therefore 
to be expected that the Poisson bracket of any constraint with 
${\cal G}^{IJ}$ or ${\cal H}_a$ vanishes on the constraint 
surface. In fact
\begin{eqnarray}
\left\{\frac{1}{2}{\cal G}^{IJ}\left[ \Lambda_{IJ}\right] ,
\frac{1}{2}{\cal G}^{KL}\left[\Omega_{KL}\right]\right\}&=&
{\cal G}^{IJ}\left[ \Lambda_{IK}\Omega_{JK}\right]\label{comga}\\
\left\{\frac{1}{2}{\cal G}^{IJ}\left[\Lambda_{IJ}\right] ,
{\cal H}_a\left[ N^a\right]\right\} &=&0\\
\left\{\frac{1}{2}{\cal G}^{IJ}\left[\Lambda_{IJ}\right] ,
{\cal H}\left[ N\right]\right\}&=&0\\
\left\{\frac{1}{2}{\cal G}^{IJ}\left[\Lambda_{IJ}\right] ,
\frac{1}{2}{\cal C}^{ab}\left[c_{ab}\right]\right\}&=&0\\
\left\{{\cal H}_a\left[ M^a\right] ,{\cal H}_b\left[ N^b\right]
\right\}&=&
{\cal H}_a\left[ M^b\partial_bN^a-N^b\partial_bM^a\right]-
\frac{1}{2}{\cal G}^{IJ}\left[ M^aN^bR_{abIJ}\right]\label{con1}\\
\left\{{\cal H}_a\left[ M^a\right] ,{\cal H}\left[ N\right]\right\}
&=&{\cal H}\left[ M^a\partial_aN-N\partial_aM^a\right] +
{\cal G}^{IJ}\left[NM^a\pi^{bIK}R_{abJK}\right]\label{con2}\\
\left\{{\cal H}_a\left[ N^a\right] ,\frac{1}{2}{\cal C}^{bc}\left[
c_{bc}\right]\right\} =
&\frac{1}{2}&{\cal C}^{ab}\left[ 
N^c\partial_cc_{ab}+2c_{ac}\partial_bN^c-c_{ab}\partial_cN^c
\right] +
\frac{\beta^2}{2(1+\beta^2)}{\cal G}^{IJ}\left[ c_{ab}N^a
\left( {^*\pi^{bIJ}}-\frac{1}{\beta}\pi^{bIJ}\right)\right]
\end{eqnarray}
The remaining Poisson brackets are
\begin{eqnarray}
\left\{{\cal H}\left[ M\right] ,{\cal H}\left[ N\right]\right\}
&=&-\frac{1}{2}{\cal H}_a\left[\left( M\partial_bN-N\partial_bM
\right)
\pi^{aIJ}\pi^b{_{IJ}}\right] +
\frac{1}{2}{\cal C}^{ab}\left[\left( M\partial_aN-
N\partial_aM\right) {^*\pi^{cIJ}}
\stackrel{(\beta)}{R}_{cbIJ}\right]
\label{hh}\\
\left\{\frac{1}{2}{\cal C}^{ab}\left[ c_{ab}\right] ,
\frac{1}{2}{\cal C}^{cd}\left[ d_{cd}\right]\right\}&=&0\\
\left\{{\cal H}\left[ N\right] ,\frac{1}{2}{\cal C}^{ab}\left[ c_{ab}
\right]\right\} &=&\frac{1}{2}{\cal D}^{ab}\left[ Nc_{ab}\right]\ ,
\label{dif}
\end{eqnarray}
where
\begin{equation}
{\cal D}^{ab}={^*\pi^{cIJ}}\left( \pi^{aIK}D_c\pi^{bJK}+
\pi^{bIK}D_c\pi^{aJK}\right) \ .\label{ncon}
\end{equation}
We note that we obtained the characteristic Poisson bracket of the 
scalar constraint in General relativity, since
\begin{equation}
-\frac{1}{2}\pi^{aIJ}\pi^b{_{IJ}}=
g(g^{tt}g^{ab}-g^{ta}g^{tb})=
q q^{ab}\ ,\label{met}
\end{equation}
where $q^{ab}$ is the 3-dimensional matrix inverse to the spatial 
part of the metric, $q_{ab}=g_{ab}$, and
\begin{equation}
q=\det(q_{ab})=\sqrt{\det\left( -\frac{1}{2}\pi^{aIJ}\pi^b{_{IJ}}
\right)}\ .
\end{equation}

Due to (\ref{dif}) the constraint algebra does not close. We shall 
not repeat the full analysis here, which is analogous to the 
standard treatment of the Hilbert-Palatini action for General 
relativity \cite{pel}. It results that the conditions 
${\cal D}^{ab}\sim 0$ must be imposed as secondary constraints, and 
it is easy to check that they form second class pairs with the 
constraints ${\cal C}^{ab}$. Their Poisson bracket is
\begin{eqnarray}
&&\left\{\frac{1}{2}{\cal C}^{ab}\left[ c_{ab}\right] ,
\frac{1}{2}{\cal D}^{cd}\left[ d_{cd}\right]\right\}=\nonumber\\
&=&{\cal C}^{ab}\left[ (c_{ab}d_{cd}-c_{ac}d_{bd}){\cal C}^{cd}+
\beta^{-1}(c_{ab}d_{cd}+c_{bd}d_{ab}-
2c_{ac}d_{bd})qq^{cd}
\right] +q^2q^{ab}q^{cd}(c_{ac}d_{bd}-c_{ab}d_{cd})\ .
\end{eqnarray}
Thus, there are no tertiary constraints, and the full Hamiltonian is
\begin{equation}
H=N{\cal H}+N^a{\cal H}_a+\frac{1}{2}\Lambda_{IJ}{\cal G}^{IJ}+
\frac{1}{2}c_{ab}{\cal C}^{ab}+\frac{1}{2}d_{ab}{\cal D}^{ab}\ .
\end{equation}

\section{Solving the second class constraints}\label{se4} 

Now one must solve both ${\cal C}^{ab}$ and ${\cal D}^{ab}$, and 
insert their solutions in the Hamiltonian, where only the scalar, 
vector and gauge constraints survive. 
The solution to ${\cal C}^{ab}$ is 
\begin{eqnarray}
\pi^{a0i}&=&E^{ai}\label{sol1}\\
\pi^{aij}&=&E^{a[i}\chi^{j]}\ ,\label{sol2}
\end{eqnarray}
where
\begin{equation}
\chi^i=-\frac{e^{ti}}{e^{tt}}\ .
\end{equation}
This is a convenient way of expressing the solution to 
${\cal C}^{ab}$, which enables us to set the time gauge in 
the simple form $\chi^i=0$. The spatial metric (\ref{met}) 
is given by 
\begin{equation}
qq^{ab}=E^{ai}\eta_{ij}E^{bj}
\end{equation}
with
\begin{equation}
\eta_{ij}=(1-\chi^k\chi^k)\delta_{ij}+\chi^i\chi^j\ .
\end{equation}
When $\chi^i\chi^i\ne 1$ this is a positive definite metric with 
the property that it does 
not distinguish between contravariant and covariant indices of 
$\chi^i$, $\eta_{ij}\chi^j=\chi^i$. The fields $E^{ai}$ are 
therefore not triads but rather they bring the metric to this 
form (they become triads after gauge fixing and 
$\eta_{ij}\to\delta_{ij}$).

Inserting the solution (\ref{sol1})-(\ref{sol2}) 
into the symplectic form (\ref{sifo}) 
projects out 12 components of the connections,
\begin{equation}
\frac{1}{2}\pi^{aIJ}\dot{\stackrel{(\beta)}{\omega}_{aIJ}}=
E^{ai}\dot A_{ai}+\zeta_i\dot\chi^i\ ,\label{conj}
\end{equation}
with
\begin{eqnarray}
A_{ai}&=&\stackrel{(\beta)}{\omega}_{a0i}+
\stackrel{(\beta)}{\omega}_{aij}\chi^j\\
\zeta_i&=&\stackrel{(\beta)}{\omega}_{aij}E^{aj}\ .
\end{eqnarray}
We are now working with the new set of canonical variables 
$(A_{ai},E^{ai})$ and $(\chi^i,\zeta_i)$ with non-trivial Poisson 
brackets
\begin{eqnarray}
\{ A_{ai}(x),E^{bj}(y)\} &=&\delta_i^j\delta_a^b\delta^3(x-y)\\
\{ \chi^i(x),\zeta_i(y)\} &=&\delta_i^j\delta^3(x-y)\ .
\end{eqnarray}
The connections can be written in terms of the new variables as 
\begin{eqnarray}
\stackrel{(\beta)}{\omega}_{a0i}&=&A_{ai}-
\stackrel{(\beta)}{\omega}_{aij}\chi^j\label{sol3}\\
\stackrel{(\beta)}{\omega}_{aij}&=&\frac{1}{2}\left( 
\frac{1}{1-\chi^k\chi^k}\epsilon_{ijk}E_{al}M^{kl}-
E_{a[i}\zeta_{j]}\right)\ ,\label{sol4}
\end{eqnarray}
where $M^{ij}$ is symmetric and represents the components of the 
connections which do not show up in the symplectic form.
Since the Poisson brackets of the vector and gauge constraints with 
${\cal C}^{ab}$ vanishes on the surface of 
primary constraints, one expects them to be straightforwardly 
written in terms of the new canonical variables. Indeed 
\begin{equation}
{\cal H}_a=E^{bi}\partial_{[a}A_{b]i}+\zeta_i\partial_a\chi^i+
\frac{\beta^2}{1+\beta^2}
\left[ -E^{b[i}\chi^{j]}A_{ai}A_{bj}-A_{ai}(\zeta_i-
\zeta_j\chi^j\chi^i)+
\frac{1}{\beta}\epsilon_{ijk}(E^{bi}A_{bj}+
\zeta_i\chi^j)A_{ak}\right] \ .\label{ga1}
\end{equation}
We split the gauge constraints into their 
boost and rotational components 
\begin{eqnarray}
{\cal G}_{boost}^i&=&{\cal G}^{0i}=\partial_a\left( E^{ai}-
\frac{1}{\beta}\epsilon_{ijk}\chi^jE^{ak}\right) -
E^{a[i}\chi^{j]}A_{aj}-\zeta_i+\zeta_j\chi^j\chi^i\label{ga2}\\
{\cal G}_{rot}^i&=&\frac{1}{2}\epsilon_{ijk}{\cal G}^{jk}=
-\partial_a\left( \epsilon_{ijk}\chi^jE^{ak}+
\frac{1}{\beta}E^{ai}\right) +
\epsilon_{ijk}(A_{aj}E^{ak}-\zeta_j\chi^k)\label{ga3}
\end{eqnarray}

In order to write the scalar constraint in terms of the new 
variables the solution to ${\cal D}_{ab}$ is required, which is 
\begin{equation}
M^{ij}=(f_{kk}-f_{kl}\chi^k\chi^l)\delta^{ij}-
(f_{kk}+f_{kl}\chi^k\chi^l)\chi^i\chi^j-
f_{ij}-f_{ji}+(f_{ik}\chi^j+f_{jk}\chi^i+f_{ki}\chi^j+
f_{kj}\chi^i)\chi^k ,\label{go1}
\end{equation}
with
\begin{equation}
f_{ij}=\epsilon_{ikl}E^{ak}\left[ 
(1+\beta^{-2})E_{bj}\partial_aE^{bl}+\chi^lA_{aj}\right]+
\beta^{-1}(E^{bk}A_{bk}\delta_{ij}-A_{bi}E^{bj}-\zeta_i\chi^j)\ .
\label{go2}
\end{equation}
Here $E_{ai}$ stands for the inverse of $E^{ai}$. 
The scalar constraint reads 
\begin{eqnarray}
{\cal H}&=&E^{ai}\chi^i{\cal H}_a+(\chi^k\chi^k-1)
\left( E^{ai}\partial_a\zeta_i+\frac{1}{2}\zeta_iE^{ai}E^{bj}
\partial_aE_{bj}\right) 
+\frac{\beta^2}{1+\beta^2}
\left\{ (\chi^k\chi^k-1)\left[ \frac{3}{4}\zeta_i\zeta_i-
\frac{3}{4}(\zeta_i\chi^i)^2-\right. \right.\nonumber\\
&&-\zeta_i\chi^iA_{aj}E^{aj}-\frac{1}{2}E^{ai}E^{bj}A_{a[i|}A_{b|j]}-
+\frac{1}{\beta}\epsilon_{ijl}
\zeta_iA_{aj}E^{al} -\frac{1}{4}(f_{ij}+f_{ji})(f_{il}+f_{li})
\chi^j\chi^l+\nonumber\\
&&\left. \left. +\frac{1}{2}f_{ij}\chi^i\chi^j
(f_{ll}+f_{lm}\chi^l\chi^m)\right] -
\frac{1}{4}\left[ (f_{ii})^2+
(f_{ij}\chi^i\chi^j)^2\right] +\frac{1}{2}f_{ij}f_{ij}\right\}
\ ,\label{ga4}
\end{eqnarray}
In this formula we dropped terms proportional to the gauge 
constraints, as we did in (\ref{ga1}). 
Therefore we 
ended up with a system described by 12 pairs of canonical 
variables $(A_{ai},E^{ai})$ and $(\chi^i,\zeta_i)$ subject to 
10 first class constraints, ${\cal G}_{boost}^i$, 
${\cal G}_{rot}^i$, ${\cal H}_a$ and ${\cal H}$. This is in 
accordance with the known result that gravity without matter 
presents 2 degrees of freedom per space point and that our theory 
is diffeomorphism invariant and possesses a further internal gauge 
symmetry, summing up to a 10 parameter family of symmetry 
transformations.

We followed this approach of introducing ${\cal C}^{ab}$ as 
primary constraints, generating secondary constraints and solving 
the resulting second class pairs, in order to keep in line with 
most of the literature in the area. But one can also insert 
(\ref{sol1})-(\ref{sol2}) and (\ref{sol3})-(\ref{sol4}) directly 
in the Lagrangian. Then expression (\ref{conj}) for the symplectic form follows and the field $M_{ij}$ shows up only in the 
scalar constraint, in a quadratic form. Variation of the 
action with respect to $M_{ij}$ leads to the solution. 
(\ref{go1})-(\ref{go2}), rendering the two methods equivalent 
(In fact there is another solution, the vanishing of the lapse 
function $N=0$, which we discard because it describes a 3 degree 
of freedom system of degenerate metrics. This same solution is 
also obtained using the method that we followed in this paper 
\cite{pel}.)

The constraint analysis ends here. The system presents its full 
gauge symmetry and it is not yet in the form given by Barbero 
\cite{bar}. In order to do so one must fix the gauge freedom for 
boost transformations. We choose the time gauge
\begin{equation}
\chi^i=0\ ,\label{gafi}
\end{equation}
and solve ${\cal G}_{boost}^i$ to obtain its canonical pair 
$\zeta_i$,
\begin{equation}
\zeta_i=\partial_aE^{ai}\ .
\end{equation}
Plugging these expressions into equations (\ref{ga1}), 
(\ref{ga2}), (\ref{ga3}) and (\ref{ga4})-(\ref{go2}) 
one arrives at Barbero's 
form of gravity. In the next section we make a shortcut to this 
derivation.

\section{Gauge fixing}\label{se5}

One can safely fix the time gauge after all secondary constraints 
have been derived, and it turns out to be simpler to set the 
gauge fixing condition (\ref{gafi}) before solving the second class 
constraints. Therefore we skip the last section and restart 
this section from the end of section \ref{seci}.

The gauge fixing condition (\ref{gafi}) together with the 
constraints ${\cal C}^{ab}$ can be written in the form
\begin{equation}
\pi^{aij}=0\ ,\label{gafi2}
\end{equation}
and only the time-space components of the $\beta$-dependent 
connection show up in the symplectic form, 
\begin{equation}
\frac{1}{2}\pi^{aIJ}\dot{\stackrel{(\beta)}{\omega}_{aIJ}}=
E^{ai}\dot A_{ai}\ ,
\end{equation}
where
\begin{eqnarray}
\stackrel{(\beta)}{\omega}_{a0i}=A_{ai}\label{ff1}\ .
\end{eqnarray}
It is convenient to write the rotational part of the 
$\beta$-dependent connection in terms of $A_{ai}$ and of the 
rotational part of the $\beta$-independent connection 
\begin{equation}
\stackrel{(\beta)}{\omega}_{aij}=\epsilon_{ijk}\left[ 
(1+\beta^{-2})\Gamma_{ak}+\alpha A_{ak}
\right]\ .\label{ff2}
\end{equation}
with
\begin{equation}
\Gamma_{ai}=\frac{1}{2}\epsilon_{ijk}\omega_{ajk}\ .\label{rot}
\end{equation}

The gauge constraints and ${\cal D}_{ab}$ become, 
using (\ref{gafi2}) and (\ref{ff1})-(\ref{ff2}), 
\begin{eqnarray}
{\cal G}_{boost}^i&=&
\partial_aE^{ai}+\epsilon_{ijk}\left[ (1+\beta^{-2})\Gamma_{aj}+
\beta^{-1}A_{aj}\right] E^{ak}\\
{\cal G}_{rot}^i&=&
-\beta^{-1}\partial_aE^{ai}+\epsilon_{ijk}A_{aj}E^{ak}\\
{\cal D}^{ab}&=&\epsilon_{ijk}E^{ci}\left[ (\partial_cE^{aj}+
\epsilon_{jmn}\Gamma_{cm}E^{an})E^{bk}+
(\partial_cE^{bj}+\epsilon_{jmn}\Gamma_{cm}E^{bn})
E^{ak}\right]\label{fu1}\ .
\end{eqnarray}
The constraints ${\cal D}^{ab}$ together with the 
following combination of the gauge constraints,
\begin{equation}
\frac{\beta}{1+\beta^2}(\beta {\cal G}_{boost}^i-{\cal G}_{rot}^i)=
\partial_aE^{ai}+\epsilon_{ijk}\Gamma_{aj}E^{ak}\ ,\label{fu2}
\end{equation}
can be solved for the variables $\Gamma_{ai}$. Equations 
(\ref{fu1})-(\ref{fu2}) are equivalent to
\begin{equation}
D_aE^{bi}=\partial_aE^{bi}+\tilde\Gamma_{ac}{^b}E^{ci}-
\tilde\Gamma_{ca}{^c}E^{bi}+\epsilon_{ijk}\Gamma_{aj}E^{bk}=0\ ,
\end{equation}
where $\tilde\Gamma_{ab}{^c}$ is the Riemannian connection 
constructed from the spatial metric (\ref{met}) 
\begin{equation}
q^{ab}=EE^{ai}E^{bi}\ ,
\end{equation}
with $E=1/\det (E^{ai})$. 
Therefore $\Gamma_{ai}$, the rotational part of the 
$\beta$-independent spin-connection, 
is nothing but the spin-connection which 
annihilates the covariant derivative of the densitized tetrad 
$E^{ai}$. That is, with $E_{ai}$ the inverse of $E^{ai}$, 
we may write the constraints 
(\ref{fu1})-(\ref{fu2}) in the form 
\begin{equation}
\Gamma_{ai}=\frac{1}{2}\epsilon_{ijk}E^{bj}\left( 
\partial_{[b}E_{a]k}+E_{a[l}E^c{_{k]}}\partial_bE_{cl}
\right)\ ,
\label{gam}
\end{equation}
which, considering the expression (\ref{rot}) for $\Gamma_{ai}$ 
in terms of the original variables $^{(\beta)}\omega_{a0i}$ and 
$^{(\beta)}\omega_{aij}$, clearly 
form second class pairs with (\ref{gafi2}). In this way we have 
proven that the gauge fixing condition chosen is independent of 
the remaining constraints, that it forms second class pairs with 
the boosts, and that it does not destroy the second class 
relation between ${\cal C}^{ab}$ and ${\cal D}^{ab}$. 

The remaining constraints are (we drop the label of the rotational 
gauge constraint)
\begin{eqnarray}
{\cal G}^i&=&-\beta^{-1}D_aE^{ai}\label{fa1}\\
{\cal H}_a&=&E^{bi}F_{abi}+
(\beta A_{ai}+\Gamma_{ai})({\cal G}_{rot}^i+\beta^{-1}
{\cal G}_{boost}^i)
\label{fa2}\\
{\cal H}&=&-\beta^{-1}\frac{1}{2}\epsilon_{ijk}E^{ai}E^{bj}\left[ 
F_{abk}+(\beta+\beta^{-1})R_{abk}\right]\ ,\label{fa3}
\end{eqnarray}
where
\begin{equation}
D_aE^{ai}=\partial_aE^{ai}-\beta\epsilon_{ijk}A_{aj}E^{ak}
\end{equation}
and $F_{abi}$ and $R_{abi}$ stand for the 
curvature of $A_{ai}$ and $\Gamma_{ai}$ respectively, 
\begin{eqnarray}
F_{abi}=\partial_{[a}A_{b]i}-\beta\epsilon_{ijk}A_{aj}A_{bk}\\
R_{abi}=\partial_{[a}\Gamma_{b]i}+
\epsilon_{ijk}\Gamma_{aj}\Gamma_{bk}\ .\label{rr}
\end{eqnarray}
We ended up with the pairs of canonical variables 
$(A_{ai},E^{ai})$ subject to the constraints 
(\ref{fa1})-(\ref{fa2})-(\ref{fa3}), which is Barbero's theory with 
coupling constant $\beta$ \cite{bar}. 

Consistency can be checked by letting $\beta\to\infty$ and 
recovering the usual formulation of General relativity with 
tetrads \cite{pel}, and setting $\alpha=\pm i$ to obtain Ashtekar's 
Hamiltonian \cite{ash}.
We end this paper with a remark on Holst's calculation \cite{hol}. 
There the author fixes the gauge (\ref{gafi}) before performing 
the Hamiltonian analysis. While this is a questionable method for 
a general gauge transformation, in this case the gauge fixing 
required the use of the gauge parameter and no time derivatives 
of it. It does not impose any conditions on the Lagrange 
multipliers and it corresponds to a so-called canonical gauge. 
Such a gauge fixing can be done directly in the Lagrangian without 
affecting locally the theory.

\bigskip
\bigskip

I thank Ingemar Bengtsson and S\"oren Holst for calling my 
attention to this problem and Marc Henneaux for a comment. 

{\it Note added:} After the completion of this work, a paper by 
S.Alexandrov \cite{ale} appeared concerning the same problem but 
with a somewhat different approach. 
To facilitate comparison I have adapted my notation a little.

\end{document}